# A Tidally Distorted Dwarf Galaxy near NGC 4449

R.M. Rich[1,4], M.L.M. Collins[2], C.M. Black[1], F.M. Longstaff[3,4], A. Koch[5], A. Benson[6], D.B. Reitzel[7,1]

[1]*Department of Physics and Astronomy, PAB 430 Portola Plaza Box 951547, UCLA, Los Angeles, CA 90095-1547, USA*

[2] *Max Planck Institute für Astronomie, Königstuhl 17, Heidelberg, D-69117, Germany*

[3]*UCLA Anderson School of Management, 110 Westwood Plaza, Los Angeles, CA 90095-1481*

[4]*Polaris Observatory Association, Frazier Park, CA 93225*

[5] *Zentrum für Astronomie der Universitat Heidelberg, Landessternwarte, Königstuhl 12, 69117 Heidelberg, Germany*

[6]*Department of Astronomy, Caltech, MC 249-1, 1200 East California Blvd, Pasadena CA 91125*

[7]*Griffith Observatory, 2800 E. Observatory Rd., Los Angeles, CA 90027*

**NGC 4449 is a nearby Magellanic irregular starburst galaxy[1] with a B-band absolute magnitude of -18 and a prominent, massive, intermediate-age nucleus[2] at a distance from Earth of 3.8 megaparsecs (ref. 3). It is wreathed in an extraordinary neutral hydrogen (H I) complex, which includes rings, shells and a counter-rotating core, spanning 90 kiloparsecs (kpc; refs 1, 4). NGC 4449 is relatively isolated[5], although an interaction with its nearest known companion—the galaxy DDO 125, some 40 kpc to the south—has been proposed as being responsible for the complexity of its HI structure[6]. Here we report the presence of a dwarf galaxy companion to NGC 4449, namely NGC 4449B. This companion has a V-band absolute magnitude of -13.4 and a half-light radius of 2.7 kpc, with a full extent of around 8 kpc. It is in a transient stage of tidal disruption, similar to that of the**

**Sagittarius dwarf[7] near the Milky Way. NGC 4449B exhibits a striking S-shaped morphology that has been predicted for disrupting galaxies[7,8] but has hitherto been seen only in a dissolving globular cluster[9]. We also detect an additional arc or disk ripple embedded in a two-component stellar halo, including a component extending twice as far as previously known, to about 20 kpc from the galaxy's centre.**

We obtained deep imaging of NGC 4449 during the time period 29 May 2011 to 1 June 2011, in the course of commissioning a 0.7-m telescope[10] designed to study low-surface-brightness structures in the vicinity of other galaxies. We discovered the profoundly tidally distorted dwarf galaxy NGC 4449B, and recover an additional lower luminosity arc or disk ripple, deeper in its halo (Fig. 1). Our photometry reveals that the original exponential halo terminates in a dumb-bell shaped shelf, beyond which we measure a de Vaucouleurs $r^{1/4}$ surface brightness profile to 20 kpc (here r is the angular distance from the centre of NGC 4449). (Figs 1 and 2). Although we do not measure a change in the g-r colour of the outer halo, the break in structure might imply a different origin for the $r^{1/4}$ component.

The lower-luminosity arc or ripple mentioned above is revealed by subtracting a model halo profile, but can also be clearly seen in unprocessed images (Fig. 1) and is also faintly visible and noted in earlier images[1]. However, we detect no additional components of a putative shell system as might be expected if this arc were part of a typical shell complex (even induced via an unusual collision geometry[11,12]). The arc or ripple might plausibly be a disk ripple, owing its origin to the interaction with NGC 4449B or a different event[13]. The ripple is 2.6 kpc long with an r-band magnitude of 19.1 (corresponding to an absolute magnitude $M_r$ = -8.91±0.1, faint even relative to Milky Way dwarfs); our halo subtraction uncovered no additional arcs or candidate dwarfs. NGC 4449B (also known as NGC 4449 J1228.8+4357.8) lies at a projected distance of 9 kpc from the nucleus of NGC 4449 at right ascension (2000) 12 h 28 min 45 s, declination (2000) 143º 57ᵈ 44", which is 39.8" E and 8' 07.2" S of the nucleus. The halo model subtraction in Fig. 1 reveals the complete extent of the dwarf, including a plume of faint emission extending northwest towards the nucleus. We derive r = 14.47 ± 0.1 mag and adopting $(m-M)_0$ = 27.91 (ref. 3), calculate $M_r$ = -13.44 ± 0.1 mag. The colour of g-r = 0.48 ± 0.22 is that of a non-star forming dwarf galaxy, consistent with the lack of structure at this location in published HI maps[1,4] and non-detection on archival GALEX[14] satellite images of 3,283 s duration in the near-ultraviolet and 1,685 s in the far-ultraviolet. It is noteworthy that the position of NGC 4449B misses by 0.3 kpc any catalogued[1] HI cloud or shell, although the position falls on the southern edge of the main HI ring[4]. Although GALEX far-ultraviolet imaging usually detects stellar emission in HI tidal tails[15], the extensive HI complex near NGC 4449 is surprisingly undetected in the GALEX imagery. DDO 125 ($M_v$=-15.57) is detected easily in HI and GALEX near- and far-ultraviolet[16], but is disjoint from the main HI complex and, lying 31 kpc to the south of NGC 4449B, is uninvolved with the dwarf. Optical emission from NGC 4449B is traceable to a full extent of 2.65' × 4.09' or 2.9 × 7.4 kpc in extent; we calculate a stellar mass[17] of 3.5 x 10⁷ solar masses. The S-shaped morphology qualitatively resembles a model[7] that places a dwarf galaxy on a highly eccentric orbit, and tracks its evolution from encounter through the close approach of the dwarf galaxy (the 'impactor') to the nucleus of the primary galaxy. Such extreme orbits are proposed for other systems: for example, And XIV has kinematics consistent with a first-encounter plunge orbit with M31 (ref. 18). NGC 4449B appears to fall somewhere between time steps 2 and 3 (as shown in figure 1 of the simulation reported in ref. 7), or 5–10 crossing times past closest approach between the dwarf and nucleus, a point in the simulation where most of the dark matter still remains bound to the dwarf. The encounter geometry that we observe for NGC 4449B is also consistent with the modeled timescale[7] over which the dwarf evolves from a compact spheroid to the 'nucleus and tails' morphology prominent in the disrupted globular cluster Palomar 5[8,9]. The width of the dwarf galaxy's central region (28" = 516 pc) constrains a length scale for the pre-encounter system even though we do not discern the nucleus. If we adopt an effective radius of ~200 pc for the pre-tidal dwarf and internal velocity dispersion $\sigma=10$ km s$^{-1}$, figure 1 of the simulation[7] gives a morphology evolution timescale $t-t_p \approx 10R_c \approx 2\times10^8$ yr, where $t-t_p$ is the time since pericentre, or closest encounter, and $R_c$ is the core radius of the dwarf. Assuming that the orbit plane is roughly perpendicular to our sightline (based on the S morphology), we find a timescale of $10^8$ yr to traverse 9 kpc at 100 km s$^{-1}$, in good agreement with the simulation timescale. We speculate that NGC 4449B is on its first encounter with NGC 4449 and experienced a close passage near the nucleus of NGC 4449. This conclusion is supported by the morphology of NGC 4449B, the plume pointing at the nucleus, and the approximate agreement with the structure and timescales of the simulation[7]. The calculated timescales would not contradict the hypothesis that the NGC 4449B encounter played a role in igniting the present epoch of star formation in NGC 4449. The simulation[7] also predicts that a morphology resembling that of NGC4449B survives only for a relatively brief interval of ~5 crossing times, or ~$10^8$ yr, which may, along with its low surface brightness, account for its uniqueness[19].

**Acknowledgements:** R.M.R. acknowledges support from the National Science Foundation. The Saturn Lodge 0.7 m telescope was funded and implemented by R.M.R. and F.L. The authors acknowledge members of the Polaris Observatory Association who maintain the observatory infrastructure and who assisted in the construction and implementation of the telescope and enclosure, and J. Riffle who designed and built the Centurion 28-inch telescope. This research has made use of the NASA/IPAC Extragalactic Database (NED) and of the Sloan Digital Sky Survey.

**Author Contributions:** R.M.R. conceived the project, obtained the data, and coordinated the activity. M.L.M.C. fit the surface photometry of NGC 4449 and NGC 4449B. C.B., F.L. and D.B.R. analysed and reduced various aspects of the dataset including the surface photometry. F.L. and R.M.R. implemented the Saturn Lodge 0.7 m telescope and detector system. A.K. provided insight on dwarf galaxies and discussion, and A.B. provided a discussion of theoretical implications.

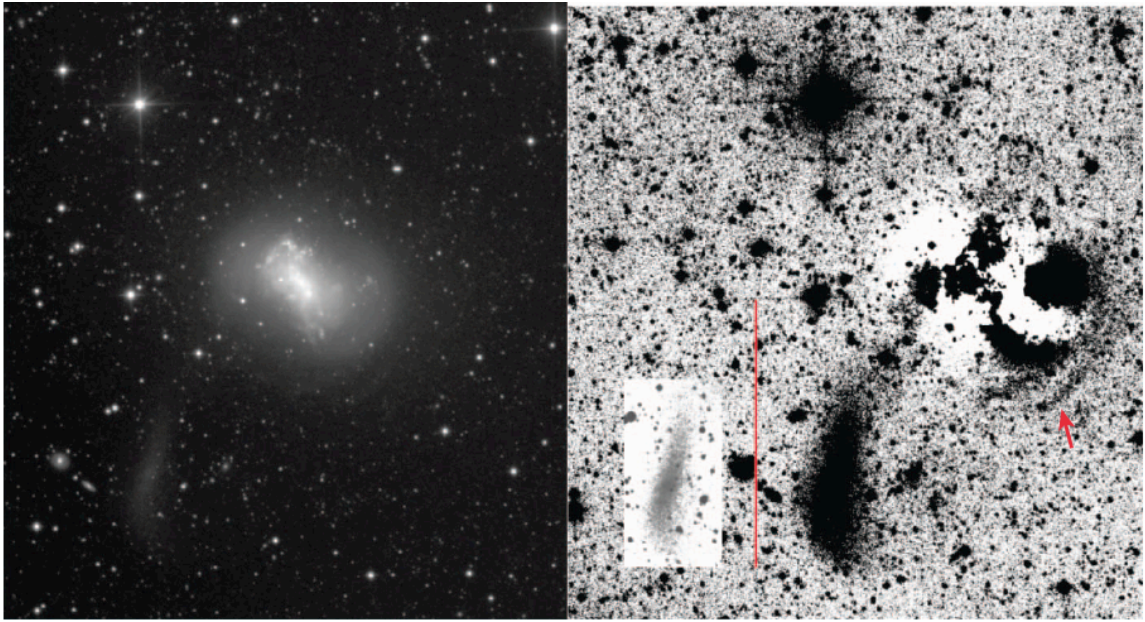

**Figure 1: Image and halo-subtracted imagery of NGC 4449.** NGC 4449 (Left) positive image of the NGC 4449 and NGC 4449B; 3.2 hr luminance filter image using an STL 11000m camera obtained using the Saturn Lodge 0.7 m Centurion[10] telescope. (Right) ELLIPSE within IRAF was used to subtract a model halo that shows detail of NGC 4449B, including a plume extended NW toward the nucleus of NGC 4449. The inset figure at left shows a softer stretch revealing the S-shape distortion characteristic of a galaxy that has undergone tidal disruption. The fainter arc/disk ripple (indicated with a red arrow) can be easily seen to the SW of the nucleus in the negative image on the left, and can be recovered as well in the positive image (right). The arc/ripple lacks the edge or counter-arc structures characteristic in classical shells. A well defined shelf in the halo of NGC 4449 is evident in the frame on the right and can be clearly seen in the surface brightness profile of Fig. 2. North up, East left; the red scale bar is 10 arcmin=11.11 kpc, adopting a distance[5] of 3.82 Mpc for NGC 4449.

Integration times were 3.12 hr in a broadband Astrodon I-series Luminance (L) filter and 45 min each in the B and R filters. The wide L filter is a square pass filter spanning 400-700nm that yields the deepest images, while the B and R filters are square pass filters covering 400-500nm and 600-700nm respectively. Because NGC 4449 is within the SDSS footprint, we use catalogued SDSS stars to calibrate B and R to SDSS g and r photometry. The total r magnitude for NGC 4449B was obtained by calibrating the L filter to SDSS r with the total magnitude from ELLIPSE, after subtracting stellar sources from the footprint of the dwarf.

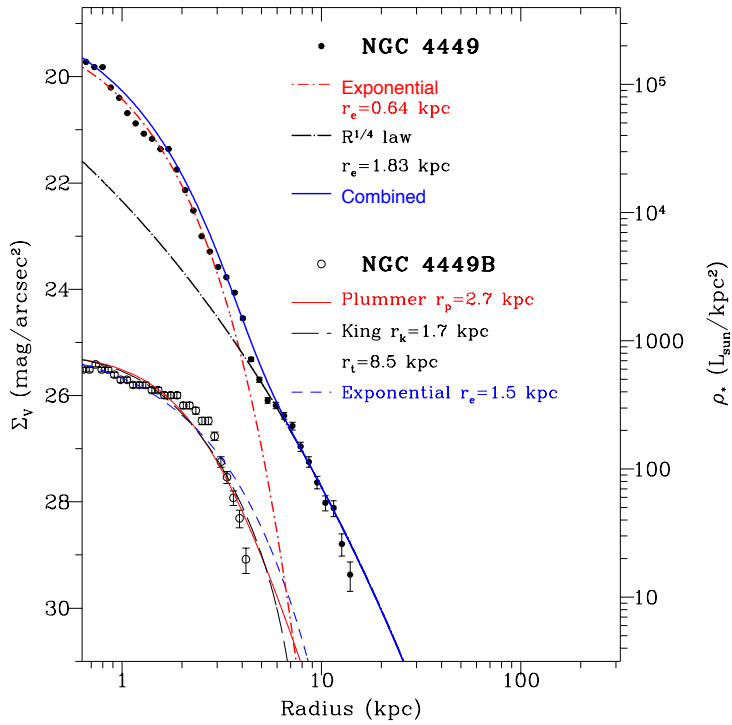

**Figure 2: Surface Photometry of NGC 4449 and NGC 4449B.** Surface brightness profiles and model fits for the halo of NGC 4449 and for NGC 4449B. The halo of NGC 4449 exhibits a dumbbell-shaped shelf (Fig 1) at 5 kpc, coincident with the break in the surface brightness profile. The inner portion of NGC 4449 is best fit by an exponential[8] with $r_e$=0.64 kpc and the outer envelope by an $r^{1/4}$ law with $r_e$=1.83 kpc that can be traced

to 20 kpc radius. Beyond 3 kpc the halo color is g-r=0.5, similar to that of the dwarf, and we do not detect any change in g-r color at the shelf; position angle and ellipticity change at the boundary of the outer halo. The exponential portion may be related to the bar, while the outer halo may have an accretion origin. NGC 4449B has a half-light radius of 2.7 kpc, but we are unable to find any analytical profile that provides a good fit, consistent with it undergoing tidal disruption[2]. Error bars are standard deviations.